\documentclass[aps,prl,twocolumn,groupedaddress,showpacs]{revtex4}

\pdfoutput=1	

\usepackage{amsmath,amsthm,amsfonts,amscd} 
\usepackage{eucal}
\usepackage{graphicx}
\usepackage{color}		
\usepackage{pxfonts}	
\usepackage{upgreek}	

\begin{document}

\title{Interference of Single Photons from Two Separate Semiconductor Quantum Dots}

\author{Edward B. Flagg}
	\email{edward.flagg@nist.gov}
\author{Andreas Muller}
\author{Sergey V. Polyakov}
\author{Alex Ling}
\author{Alan Migdall}
\author{Glenn S. Solomon}
	\email{glenn.solomon@nist.gov}
	\affiliation{Joint Quantum Institute, National Institute of Standards and Technology 
				\& University of Maryland, Gaithersburg, MD, USA.}

\date{\today}

\begin{abstract}
We demonstrate and characterize interference between discrete photons emitted by two separate semiconductor quantum dot states in different samples excited by a pulsed laser.  Their energies are tuned into resonance using strain.  The photons have a total coalescence probability of 18.1\% and the coincidence rate is below the classical limit.  Post-selection of coincidences within a narrow time window increases the coalescence probability to 47\%.  The probabilities are reduced from unity because of dephasing and the postselection value is also reduced by the detector time response.
\end{abstract}

\pacs{78.67.Hc,42.50.Ar}

\maketitle

When two classical optical fields interfere on a beamsplitter, the joint probability of detection at the two outputs can be as low as 50\% \cite{Loudon}.  In seminal work, Hong \textit{et al.\!} \cite{Hong1987PRL} showed that pairs of photons produced by parametric down-conversion (PDC) and which interfere on a beam splitter can have a reduction in coincidence detection well below 50\%, reaching zero for an ideal source of indistinguishable photon pairs.  Such interference, where two single-photon Fock states coalesce into a two-photon Fock state, has become one of the central measurements in quantum optics.  The coalescence probability hinges on the indistinguishability of the photons involved. 

Photon pairs produced by PDC are highly indistinguishable, but the number of pairs produced per pulse is super-Poissonian \cite{WallsMilburn}.  A sub-Poissonian source of indistinguishable single photons, however, is a fundamental component of emerging concepts in quantum information.  For example, such sources could be used to realize a quantum C-NOT gate using linear optical elements \cite{Knill2001Nat}.  Quantum emitters, such as trapped atoms \cite{McKeever2004Sci,Darquie2005Sci}, ions \cite{Maunz2007NatPhys}, and quantum dots (QDs) \cite{Michler2000Sci,Santori2001PRL}, are good sources of single photons.  Single photons with a high degree of indistinguishability have been produced by a single QD in an optical microcavity \cite{Santori2002Nat} while separate sources of mutually indistinguishfable photons have been produced by pairs of trapped atoms \cite{Beugnon2006Nat} and ions \cite{Maunz2007NatPhys}.  Lifetime-limited, spectrally identical photons have been produced by two separate molecules \cite{Lettow2007OE} but indistinguishability has not yet been shown.  Interference between separate solid-state photon sources, like QDs or impurities in crystals, has been performed.  The impurity case \cite{Sanaka2009PRL} beats the classical limit after subtracting a fitted background, while the QD case \cite{Benyoucef2009APL} shows no interference due to dephasing.  Photons from a single QD excited by a continuous-wave (cw) laser have shown interference which beats the classical limit in a narrow time window \cite{Patel2008PRL,Ates2009PRL}, but further evaluation is necessary to extract the two-photon coalescence.

In this Letter we demonstrate the interference of photons emitted by two semiconductor QD states, each in a different sample.  We observe a clear signature of coalescence in the second-order correlation and the coincidence detection probability is below the classical limit.  We tune the QD states into resonance using externally applied strain.  Using pulsed excitation allows us to measure the total coalescence probability of photons whose arrival time is controlled to within limits intrinsic to the QDs.  The probability of coalescence is reduced from unity mainly because of dephasing of the QD states.  The data are matched well by a model of two-photon interference using measured values from the emission of each QD.

\begin{figure*}[t]
	\includegraphics[width=3in,angle=0]{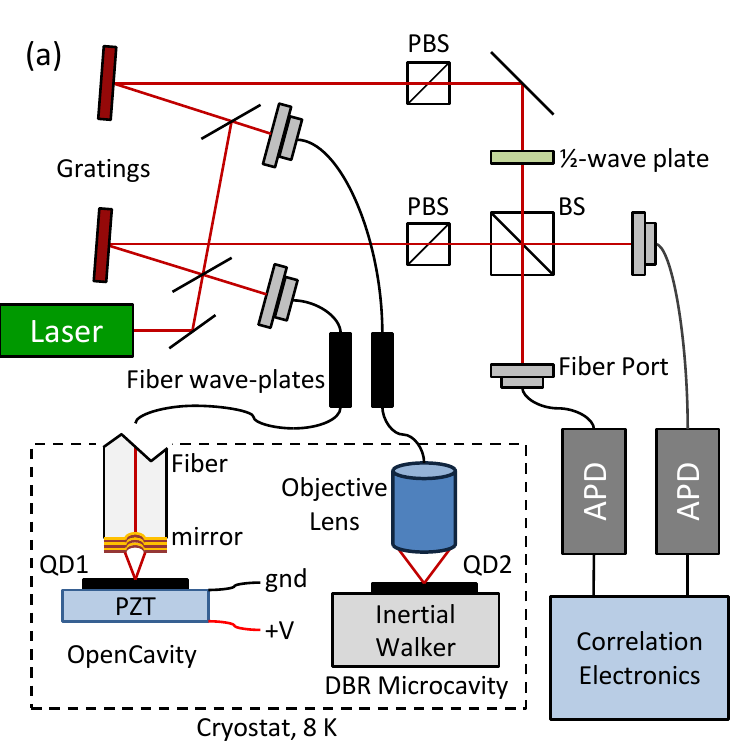}
	\includegraphics[width=4in,angle=0]{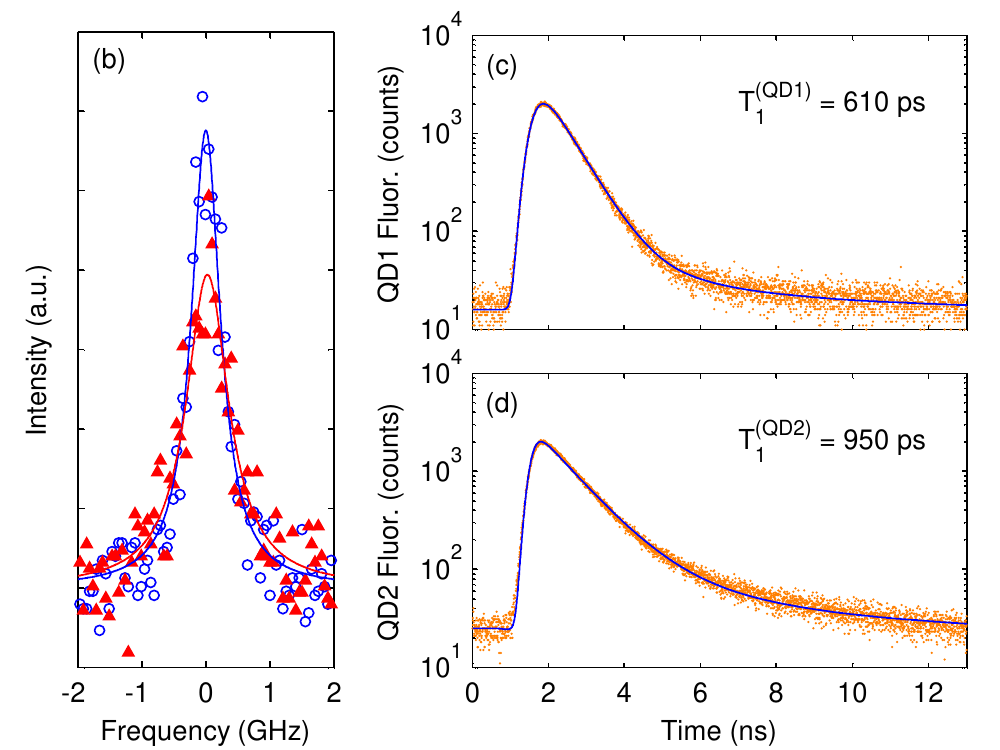}
	\caption{(Color online) (a) Schematic of QD samples and interferometer.  
	(b) Area-normalized emission lines for QD1 (\textcolor{blue}{$\circ$})
	and QD2 (\textcolor{red}{$\blacktriangle$}).
	(c) Time-dependent fluorescence from QD1.  
	(d) Time-dependent fluorescence from QD2.}
	\label{f:Fig1}
\end{figure*}

The two QD single-photon sources are in separate samples.  The samples were made using molecular-beam epitaxy and contain a low density (approximately 10 $\upmu$m$^{-2}$) of strain-induced InAs QDs.  One sample is a 4-$\uplambda$ planar distributed Bragg reflector (DBR) microcavity with 15.5 lower (10 upper) DBR pairs of GaAs and AlAs; the cavity mode is centered at $\uplambda$ = 920 nm.  The other sample is an open cavity comprising a lower DBR (35.5 pairs) and an upper external mirror attached to an optical fiber which we recently described in Ref.\! \cite{Muller2009APL}.  Figure \ref{f:Fig1}(a) shows a schematic of the two QD samples and the interferometer.  The open cavity sample is glued to a piezoelectric transducer (PZT) such that changing the voltage applied to the PZT strains the sample laterally and tunes the emission energy of the QDs \cite{Seidl2006APL}.  Both samples are maintained at 8 K in a cryostat.  The QDs are excited by a mode-locked Ti:sapphire laser with a wavelength of 800 nm and a repetition rate of 76.1 MHz (period = 13.14 ns).  After finding a QD in the open cavity which demonstrated high-quality antibunching and a narrow linewidth, denoted  QD1, we scan the DBR microcavity for a second QD, denoted QD2, whose emission energy is within the approximately 10 GHz tuning range of QD1's energy.  The emission from both QDs is coupled into optical fibers, and variable fiber wave plates ensure proper polarization matching.  The light exiting the fibers is filtered by diffraction gratings and sent to interfere at a nonpolarizing beamsplitter.  The light from the beamsplitter outputs is coupled into fibers and guided to a pair of avalanche photodiodes (APDs) with a time resolution of 640 ps.

We confirm the spectral overlap of the two QDs and measure their emission linewidths using a scanning Fabry-Perot cavity, monochromator and photodetector with an overall resolution of 150 MHz.  Figure \ref{f:Fig1}(b) shows the emission data for each QD fitted with a Lorentzian.  The excitation laser powers for each QD are adjusted such that their emission intensities are the same; thus the areas under both curves are equal.  The full-widths at half-maximum are $\Delta\nu_{\mathrm{QD1}} = 0.55 \pm 0.02$ GHz and $\Delta\nu_{\mathrm{QD2}} = 0.81 \pm 0.05$ GHz from which we extract the coherence times $T_{2}^{(\mathrm{QD1})} = 580 \pm 20$ ps and $T_{2}^{(\mathrm{QD2})} = 390 \pm 20$ ps.  From polarization-dependent measurements (not shown) we determine that the emission line from QD1 is a trion and that from QD2 is one of the fine-structure split lines of an exciton.

To measure the QD lifetimes, each QD's emission is individually sent through a monochromator to an APD and the resulting population decay curves are shown in Figs.\! \ref{f:Fig1}(c) and \ref{f:Fig1}(d) on a log scale.  The curve for QD1 is fit with a single exponential decay, while that for QD2 is fit with a biexponential because it is a neutral exciton and we must account for spin flip transitions from dark states \cite{Dalgarno2005PSSA}.  Both curves include the effect of the detector time resolution.  The lifetimes are $T^{(\mathrm{QD1})}_{1} = 610 \pm 5$ ps and $T^{(\mathrm{QD2})}_{1} = 950 \pm 5$ ps; the dark state spin flip time for QD2 is 4.0 $\pm$ 0.5 ns.  For both QDs, $T_{2} < 2 T_{1}$; i.e., the coherence times are not lifetime limited.

\begin{figure}[b]
	\vspace{-10pt}
	\includegraphics[width=3.38in,angle=0]{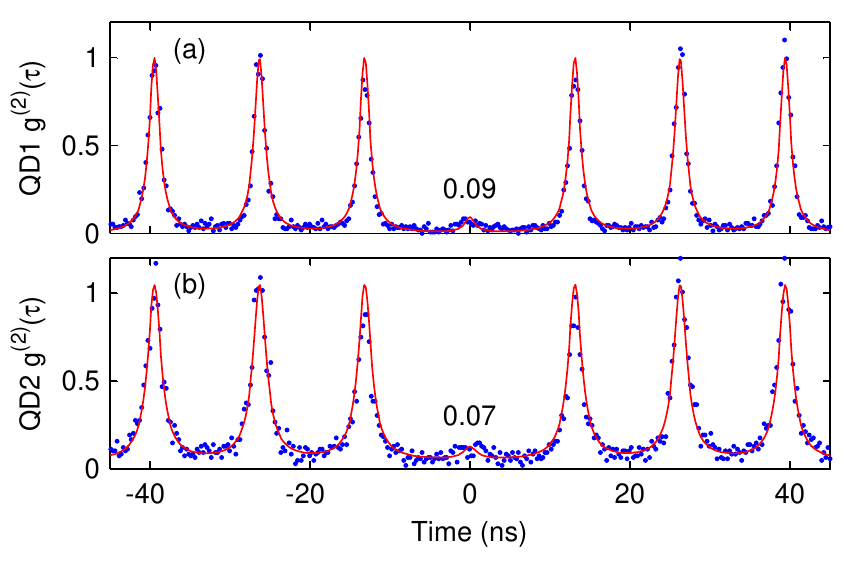}
	\caption{(Color online) (a) Autocorrelation for QD1; residual counts in the center peak are 9\% of those in the other peaks.
		(b) Autocorrelation for QD2; residual counts in the center peak are 7\%.}
	\label{f:Fig2}
\end{figure}

The beamsplitter and APDs can be used as a Hanbury Brown-Twiss correlation measurement \cite{HBT1956Nat} if we input the emission from only one of the QDs.  Figures \ref{f:Fig2}(a) and \ref{f:Fig2}(b) show the second-order autocorrelation measurement, $g^{(2)}(\tau)$, for each QD individually.  The residual counts in the $\tau=0$ peaks are 9\% and 7\% of the average amount in the other peaks for QD1 and QD2, respectively.  No background subtraction has been applied and the lack of coincidences shows that the light from each QD is over 90\% pure single-photon emission.

When photons from each QD interfere, the quality of two-photon interference depends on many experimental parameters, not only spectral overlap.  To maximize the spatial overlap of the interferometer input modes at the beamsplitter, we send cw laser light tuned to the wavelength of the QDs into both input ports.  The field intensity at the output ports can be easily visualized with a CCD camera and the spatial mode overlap optimized.  The spatial mode overlap is 95 $\pm$ 5\% as determined from the interference fringes of the laser light.  By sending the emission from each QD separately through the interferometer and using time-resolved detection, we also measure and eliminate the time delay difference between the two light collection paths to ensure optimal temporal overlap of the photons.  We attain optimal polarization alignment using the variable fiber wave plates on each input port of the interferometer and polarizing beamsplitters to highly attenuate the remaining undesired polarization.  Photons in the two beam paths can be made distinguishable by rotating a $1/2$-wave plate to make their polarizations orthogonal.  At orthogonal polarization the light does not interfere but since the inputs are single photons, coincidence detection is still below that of Poissonian light.

\begin{figure}[b]
	\includegraphics[width=3.38in,angle=0]{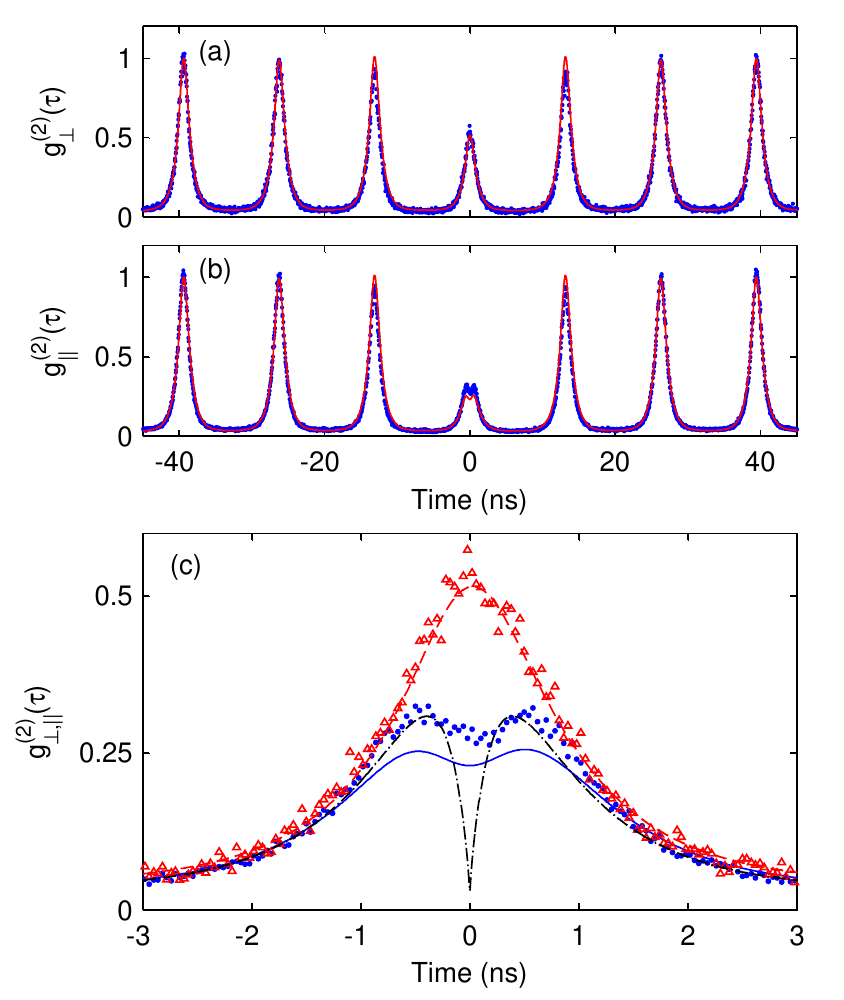}
	\caption{(Color online) (a) Correlation of the interference for orthogonal polarizations with simulated curve.
		(b) Correlation of the interference for parallel polarizations with simulated curve.
		(c) Close-up of $\tau=0$ peak for orthogonal (\textcolor{red}{$\vartriangle$}) and parallel (\textcolor{blue}{$\bullet$})  polarizations.  The solid and dashed curves are simulations including the detectors' time response, while the dash-dotted curve is the expected curve for infinitely fast detectors.}  
	\label{f:Fig3}
\end{figure}

Despite the non-negligible differences between the QDs in coherence time, lifetime, and charge state, their photons still interfere.  Figures \ref{f:Fig3}(a) and \ref{f:Fig3}(b) show the second-order correlation of the light exiting the two output ports of the interferometer for orthogonal and parallel polarizations, respectively.  No background or dark count subtraction is performed on the data.  For parallel polarizations, the height of the $\tau=0$ peak is lower than that for orthogonal polarizations, indicating that photons from the two different QDs have a nonzero coalescence probability.

An interesting feature of pulsed correlation, which is present in cw but whose significance is obscured, is the reduction in coincidences in the center of the $\tau=0$ peak for parallel polarizations.  Figure \ref{f:Fig3}(c) shows a close-up of the peak for both relative polarizations.  While the sum of coincidence counts in the $\tau=0$ peak is not influenced by the time response of the detectors, the depth of the dip is affected.  The dash-dotted curve in Fig.\! \ref{f:Fig3}(c) is the result of a simulation based on the work in Ref.\! \cite{Kiraz2004PRA} using the $T_{1}$ and $T_{2}$ values measured above.  It shows the shape expected if the detectors were infinitely fast and the single-QD $g^{(2)}(\tau)$ went to zero at $\tau=0$.  The solid curve is the same simulation including the effects of the detector response.  The residual difference between the data and the solid curve is due to the remaining imperfections in the photon overlap which are not accounted for in the simulation.  The dashed curve is a simulation of the orthogonal situation including the detector response.  It matches the data very well because the orthogonal polarizations do not interfere, and therefore the data do not depend on the photon overlap.

The probability of coalescence is given by
\begin{equation}
P_{c} = \frac{A_{\perp} - A_{\parallel}}{A_{\perp}}
\end{equation}
where $A_{\perp,\parallel}$ is the integrated number of counts in $g^{(2)}_{\perp,\parallel}(\tau)$ during one repetition period around $\tau=0$.  From the data in Fig.\! \ref{f:Fig3} we obtain $P_{c} = 18.1 \pm 0.4$\% \cite{DarkcountNote}.  This value is reduced from unity mainly due to the presence of dephasing \cite{Bylander2003} as explained below.  It would be erroneous to calculate $P_{c}$ using the values of $g^{(2)}_{\perp,\parallel}(\tau)$ right at $\tau=0$ because any photon emitted by a QD will have a nonzero temporal extent.  Using the values at $\tau=0$ results in a postselective measurement of coalescence
\begin{equation}
P'_{c} = \frac{ g^{(2)}_{\perp}(0) - g^{(2)}_{\parallel}(0) } { g^{(2)}_{\perp}(0) }
\end{equation}
which represents the probability of coalescence conditional on detecting both photons within a narrow time window.  A cw interference measurement can obtain the postselective value \cite{Patel2008PRL}, $P'_{c}$, which has been shown to be quite high for a QD in a microcavity \cite{Ates2009PRL}.  The data in Fig. \ref{f:Fig3}(c) lead to the value $P'_{c} = 47 \pm 6$\%.  With infinitely fast detectors the value of $g^{(2)}_{\parallel}(0)$ would go nearly to zero as does the dash-dotted curve in Fig.\! \ref{f:Fig3}(c).  The postselective probability would then be $P'_{c} = 96 \pm 4$\% despite the significant counts remaining in the peak at non-zero $\tau$.  

The coincidence detection rate is given by $A_{\perp,\parallel} / B$ where $B$ is the average number of integrated counts in the peaks not centered at $\tau=0$.  The classical limit is $A_{\parallel}/B = 0.5$, which is the lowest coincidence rate for two classical fields \cite{Loudon,Ou1989JOSAB}.  From the data in Fig.\! \ref{f:Fig3} we extract the value $A_{\parallel}/B = 0.481 \pm 0.002$, which is below the classical limit within experimental error.  A previous interference measurement on separate solid-state photon sources has demonstrated coincidence rates below the classical limit after postprocessing removal of fitted background coincidences \cite{Sanaka2009PRL}.  To our knowledge, the present measurement is the first done on separate solid-state photon sources which demonstrates a coincidence rate below the classical limit in the raw data.

The central dip in Fig.\! \ref{f:Fig3}(c) is caused by coalescence of the photons.  It does not completely eliminate the $\tau=0$ peak mainly because the QDs' coherence time is not lifetime limited.  Though the photons' temporal extent is given by the QD lifetimes, $T_{1}$, the time delay over which they can interfere is given by the coherence times, $T_{2}$.  Thus the width of the peaks are determined by $T_{1}$, and the width of the dip is determined by $T_{2}$.  If the coherence times were lifetime-limited we would have $T_{2} = 2 T_{1}$ and the dip would be wide enough to nearly eliminate the $\tau=0$ peak.  Some residual counts would remain because the two QD lifetimes are different.  When $T_{2} < 2 T_{1}$ as in this case, the dip is narrow enough to leave significant counts in the peak and gets almost smoothed away by the finite time response of the detectors.  Thus for a postselective measurement, both the time window and the detector response time must be less than $T_{2}$ \cite{Patel2008PRL}.  It is important to note that the imperfect coalescence in the present work is not caused mainly by the differences in the two QDs' lifetimes and coherence times, but by the fact that the coherence times are not lifetime-limited.

For photons from two QDs with the values of $T_{1}$ and $T_{2}$ we measure, the maximum theoretical coalescence probability is $P_{c,\mathrm{max}} = 29$\%, as determined by the simulations described above.  We attribute the difference between $P_{c,\mathrm{max}}$ and our measured value of 18.1\% to dark counts and background scattering, which show up as a constant offset in the correlation functions, and to remaining imperfect photon overlap.

The indistinguishability of two photons is a property independent of the measuring device. Therefore, while a cw photon source could be utilized in an application requiring indistinguishable photons, postselection will be required if controlling the photon arrival time is necessary.  However, to accurately characterize the indistinguishability an unconditional measurement must be performed. This will be facilitated in the future with pulsed excitation sources of indistinguishable photons.

We acknowledge partial support from the NSF Physics Frontier Center at the Joint Quantum Institute.

\textit{Note added in proof.} -- We have recently become aware of similar work that is under review \cite{Patel2009arXiv}.

\bibliography{HOM}

\begin{thebibliography}{25}
\bibitem[{Dar()}]{DarkcountNote}
\bibinfo{note}{Removing the effect of dark counts on the data raises $P_{c}$ to
  19.7 $\pm$ 0.4 \%. Dark counts are uncorrelated with other counts, so their
  effect can be determined from the correlation of detector signals
  $(I_{1}+D_{1})\otimes(I_{2}+D_{2})$ where $I_{i}$ is the signal from photons,
  and $D_{i}$ is the dark counts.}
\end{thebibliography}

\end{document}